\documentclass[conference]{IEEEtran}
\IEEEoverridecommandlockouts

\usepackage{cite}
\usepackage{amsmath,amssymb,amsfonts}
\usepackage{algorithmic}
\usepackage{graphicx}
\usepackage{textcomp}
\usepackage{xcolor}
\usepackage{booktabs}
\usepackage{graphicx} 
\usepackage{comment}

\def\BibTeX{{\rm B\kern-.05em{\sc i\kern-.025em b}\kern-.08em
    T\kern-.1667em\lower.7ex\hbox{E}\kern-.125emX}}
\begin{document}

\title{“Everyone Else Does It”: The Rise of Preprinting Culture in
Computing Disciplines\\
}

\author{
\centering
\begin{tabular}{ccc}
\begin{minipage}[t]{0.31\textwidth}
\centering
Kyrie Zhixuan Zhou\\
\textit{University of Texas at San Antonio}\\
kyrie.zhou@utsa.edu
\end{minipage} &
\begin{minipage}[t]{0.36\textwidth}
\centering
Justin Eric Chen\\
\textit{University of Illinois Urbana-Champaign}\\
jechen4@illinois.edu
\end{minipage} &
\begin{minipage}[t]{0.3\textwidth}
\centering
Xiang Zheng\\
\textit{University of Wisconsin-Madison}\\
xzheng246@wisc.edu
\end{minipage} \\[1.2em]
\\[0.5em]
\begin{minipage}[t]{0.3\textwidth}
\centering
Yaoyao Qian\\
\textit{Northeastern University}\\
qian.ya@northeastern.edu
\end{minipage} &
\begin{minipage}[t]{0.3\textwidth}
\centering
Yunpeng Xiao\\
\textit{Emory University}\\
yunpeng.xiao@emory.edu
\end{minipage} &
\begin{minipage}[t]{0.3\textwidth}
\centering
Kai Shu\\
\textit{Emory University}\\
kai.shu@emory.edu
\end{minipage}
\end{tabular}
}


\maketitle

\begin{abstract}
  Preprinting has become a norm in fast-paced computing fields such as artificial intelligence (AI) and human-computer interaction (HCI). In this paper, we conducted semi-structured interviews with 15 academics in these fields to reveal their motivations and perceptions of preprinting.
    The results found a close relationship between preprinting and characteristics of the fields, including the huge number of papers, competitiveness in career advancement, prevalence of scooping, and imperfect peer review system -- preprinting comes to the rescue in one way or another for the participants.
    Based on the results, we reflect on the role of preprinting in subverting the traditional publication mode and outline possibilities of a better publication ecosystem. 
    Our study contributes by inspecting the community aspects of preprinting practices through talking to academics.
\end{abstract}
\begin{IEEEkeywords}
Preprint, Culture, Computer Science, AI, HCI 
\end{IEEEkeywords}

\section{Introduction}

A preprint is a scholarly manuscript that authors post to an open-access repository or platform to facilitate the rapid and unrestricted dissemination of their research \cite{bourneTenSimpleRules2017, pueblaPreprintsTheirEvolving2022}. 
Unlike most journals or conferences, preprint servers do not require peer review prior to publication. 
Instead, they conduct only basic screening (e.g., for plagiarism or offensive content) and typically post submissions within a few days \cite{hoyRiseRxivsHow2020}. 
Authors retain copyright and can choose open licenses that facilitate the reuse of their work. 
With the preprint model, authors can control the dissemination process of their work as preprints may be posted before, during, or independently of submission \cite{chiarelliPreprintsScholarlyCommunication2019, fraserMotivationsConcernsSelection2022}. 
Some scholars call preprints “e-print”, given that some ``preprints" are published in peer-reviewed venues \cite{yudhoatmojoUnderstandingUseEPrints2023}. 
These repositories generally preserve submissions permanently and, when applicable, provide links to the final published versions.

In addition to research articles, preprint repositories can host a wide range of scholarly outputs, including postprints, conference papers, working papers, technical reports, literature reviews, book chapters, slide decks, and posters \cite{pueblaPreprintsTheirEvolving2022}. 
For research articles, if later published, the content of a preprint typically aligns with that of a manuscript submitted to a scholarly journal or conference, with most subsequent revisions being minor \cite{brierleyTrackingChangesPreprint2022, kleinComparingPublishedScientific2019}. 
This breadth of document types and the free access have expanded the role of preprint servers from mere archives to comprehensive hubs of scholarly communication, offering rapid visibility and enabling early community feedback long before formal publication.

Due to preprints’ various advantages, preprints have rapidly increased their popularity in recent years. 
The adoption of preprints has since been strongest in physics and astronomy, followed by mathematics, computer science, and quantitative biology \cite{dorta-gonzalezEffectPerceivedPreprint2025, niPreprintNotPreprint2023}. 
Fields like biomedical science also saw a marked surge in preprint posting in response to the COVID-19 pandemic, driven by the urgent need for timely access to empirical results \cite{gianolaCharacteristicsAcademicPublications2020, zhengSignificantShorttermInfluence2024}. 

As for computing fields, usage of arXiv has risen dramatically among the top conferences in computer science \cite{suttonPopularityArXivOrg2017}. 
Lin et al. estimated that approximately 77\% of computer science preprints on arXiv were eventually published in peer-reviewed venues, underscoring the close relationship between preprint posting and formal publication \cite{linHowManyPreprints2020}. 
Furthermore, since ChatGPT’s advent in 2022, these fields, especially AI-related domains, have seen an unprecedented surge in preprints \cite{adarkwahArePreprintsThreat2025}.

This study aims to understand the interaction between the rise of preprinting culture and the intrinsic characteristics of computing disciplines such as computer science and information science, as well as some particular research fields such as artificial intelligence (AI) and human-computer interaction (HCI).
These characteristics include the increase in both paper numbers and researchers \cite{azad2024publication}, competitiveness of graduate school applications and faculty positions, and fast evolution of technology \cite{penalvo2023we}.
As a result, academics in these fields may find it more necessary to preprint their research in order to ensure timeliness and novelty, or achieve their career goals.

To understand researchers' motivations and perceived pros and cons of preprinting in competitive fields such as AI and HCI, we conducted semi-structured interviews with researchers in different career stages.
We aimed to answer three research questions with the current study:
\begin{itemize}
    \item \textbf{RQ1:} How do researchers practice preprinting in AI and HCI?
    \item \textbf{RQ2:} How are preprinting practices influenced by computing culture?
    \item \textbf{RQ3:} How do researchers perceive the pros and cons of preprinting in AI and HCI?
\end{itemize}

Our research makes two major contributions.
First, we are one of the first to qualitatively understand researchers' perceptions of preprinting in competitive computing fields.
Second, by investigating motivations, as well as pros and cons of preprinting, we aspire to provide insights into better publication modes in the future.

\section{Related Work}

The growing adoption of preprints has generated extensive discussion in existing studies regarding their benefits and drawbacks. 
In this section, we discuss literature on the benefits and drawbacks of preprints and publication culture in computing disciplines.

\subsection{Benefits of Preprints}

Proponents of preprints argued that by embodying the principles of open science, which highlight transparency, inclusivity, rapid sharing, and equal access, preprints contribute to a more open and efficient scholarly ecosystem \cite{besanconOpenScienceSaves2021, ginspargPreprintDejaVu2016, sarabipourValuePreprintsEarly2019, smartEvolutionBenefitsChallenges2022}. 
For authors, one appeal of preprints is that preprints enable authors to share their findings openly and promptly, bypassing the lengthy editor evaluation and peer-review cycle \cite{bourneTenSimpleRules2017, fraserMotivationsConcernsSelection2022}. 
Early sharing increases visibility and public awareness of one’s work, facilitating timely dissemination, feedback, and collaboration across the scientific community \cite{fraserMotivationsConcernsSelection2022}.

Since preprints are date-stamped and permanent citable records, they provide clear records of productivity that scholars can share during hiring, promotion, and grant evaluations \cite{sarabipourValuePreprintsEarly2019}. 
In fields where “negative” or null results are not favored by editors and peer reviewers and thus difficult to publish, preprints offer an essential venue for disseminating all scientifically valid outcomes \cite{bourneTenSimpleRules2017, ginspargLessonsArXivs302021}. 
This feature of inclusiveness accelerates the communication of all types of discoveries, reducing duplication of effort and guiding researchers away from unproductive avenues \cite{pueblaPreprintsTheirEvolving2022}.

Crucially, preprints establish priority of discovery, protecting authors from “being scooped” and incentivizing rigorous work, with the unalterable date stamps being widely recognized across disciplines as proof of precedence \cite{chiarelliPreprintsScholarlyCommunication2019, ginspargPreprintDejaVu2016, hillScoopedEstimatingRewards2025}. 
This is a major priority in many fast-moving disciplines \cite{chiarelliPreprintsScholarlyCommunication2019}. 
Moreover, because preprints are openly available, authors often expect increased visibility, reflected by downloads, social media mentions, and early citations to their work \cite{fraserMotivationsConcernsSelection2022, kleinComparingPublishedScientific2019}. 
Empirical evidence shows that journal articles first shared as preprints tend to receive higher citation and Altmetric counts than those that were not preprinted \cite{fraserRelationshipBioRxivPreprints2020, fuReleasingPreprintAssociated2019}. 
These benefits may contribute to authors’ motivation to publish preprints before submitting elsewhere.

\subsection{Drawbacks of Preprints}

Despite these benefits, several concerns about preprints still exist. 
The absence of a peer review process for preprints may lead readers to question the credibility and trustworthiness of preprinted work \cite{adarkwahArePreprintsThreat2025, niPreprintNotPreprint2023, soderbergCredibilityPreprintsInterdisciplinary2020}. 
Surveys indicate that many researchers still view peer review as the gold standard for assessing quality and are hesitant to cite or use preprints in evaluation contexts \cite{nicholasEarlyCareerResearchers2025, tenopirTrustworthinessAuthorityScholarly2016}. 
Compared with journal or conference publications, assessing the quality of preprints by signals such as venues’ reputation and peer review comments is more challenging \cite{adarkwahArePreprintsThreat2025}. 
Without robust quality assurance mechanisms, the relentless daily influx of preprints can overwhelm readers, leading to information overload and diluting attention \cite{smartEvolutionBenefitsChallenges2022}.

Another major concern is the potential for premature media coverage. 
Sharing results before peer review can heighten the risk of disseminating unverified or erroneous findings, potentially contributing to misinformation \cite{niPreprintNotPreprint2023}.
The research findings from preprints may also be misinterpreted by online users, whether intentionally or not, to support their narratives \cite{yudhoatmojoUnderstandingUseEPrints2023}. 
Without the gatekeeping role of peer review, this concern becomes particularly acute when findings receive media attention and are translated into real-world practices \cite{zhengDissectingGlobalPeer2025, zhengEffectivenessPeerReview2023}. 

Moreover, preprints may potentially conflict with journal or conference policies. 
Some journals still consider preprints as ``prior publication,'' which may affect the acceptance of the manuscript \cite{nallamothuPreprintsCardiovascularScience2017}. 
Although many publishers have moved away from this convention, it persists or is not clearly disclosed in certain venues \cite{chaleplioglouPreprintPaperPlatforms2023, klebelPeerReviewPreprint2020, teixeiradasilvaPreprintPolicies142019}.
Preprints may also compromise the anonymity central to double-blind peer review. 
The public availability can reveal the authors’ identities, which may influence reviewers based on author prestige or other characteristics \cite{fatoneChallengesDoubleblindPeer2020}.
Preprints may also influence editorial or peer-review judgments through existing public discussions and feedback before formal submission \cite{chaleplioglouPreprintPaperPlatforms2023}.
Analyzing 5,027 submissions to a major computer science conference, Sun et al. provided empirical support for the effectiveness of double-blind review in mitigating biases \cite{sunDoesDoubleblindPeer2022}. 
However, preprints break the anonymity, as Rastogi et al. found that over one-third of reviewers at two top computer science conferences admitted to searching online for assigned papers \cite{rastogiArXivNotArXiv2022}. 
Bharadhwaj et al. showed that less confident reviewers were more likely to give higher scores to papers with well-known authors at a top computer science conference \cite{bharadhwajDeanonymizationAuthorsArXiv2020}. 

A common myth about preprints is the risk of being scooped: if research findings are shared as preprints, others might be first to publish those ideas or results, thereby denying authors the recognition they deserve. 
However, researchers’ perceptions of this risk may not align with reality \cite{pueblaPreprintsTheirEvolving2022}. 
In fact, there is little evidence that preprints lead to scooping or encourage it \cite{ngInternationalCrosssectionalSurvey2024}. 
On the contrary, as mentioned above, preprint servers help protect against scooping by time-stamping submissions and establishing priority of discovery \cite{chiarelliPreprintsScholarlyCommunication2019, ginspargPreprintDejaVu2016}.

\subsection{Publication Culture in Computing}

Computing is a young yet fast-evolving discipline \cite{franceschetRoleConferencePublications2010}. 
In recent decades, many rising topics such as AI, HCI, and large language models (LLMs) have garnered tremendous attention and are being applied across disciplines and industries.
Universities and technology companies are investing heavily in computing techniques, such as machine learning, to enhance prediction, analysis, and communication \cite{sharmaMachineLearningDeep2021}. 
As Suleyman characterizes, the “coming wave” of technologies centered on AI will bring transformative applications that promise to both empower humanity and introduce unprecedented risks \cite{suleymanComingWaveAI2023}. 
With a large and rapidly expanding researcher community, computing is advancing at such a breakneck pace that many feel anxious about keeping up with continual breakthroughs \cite{togeliusChooseYourWeapon2024}. 
Consequently, the pressure to pursue the most recent findings and technology has become a defining feature of computing today, which may stimulate the usage of preprints that facilitate immediate publishing and fast dissemination.

Moreover, unlike most academic fields, traditionally computer science relies predominantly on conference publications, with program committees shouldering the bulk of peer review \cite{franceschetRoleConferencePublications2010, leeToxicCultureRejection2022}. 
The “publish-or-perish” culture of academia and the exceptionally low acceptance rates at top-tier conferences exacerbate a deluge of (re)submissions, overwhelming reviewers and organizers \cite{carsonOverCompetitivenessAcademiaLiterature2013, parhamiLowAcceptanceRates2016}. 
With novelty and originality being the primary criteria for acceptance, scholars need to demonstrate their findings to be state-of-the-art, and they are the first to propose them \cite{leeToxicCultureRejection2022}. 
Therefore, the imperative to demonstrate innovation and secure acceptance at top conferences has fostered a hyper-competitive environment, shaping research agendas around incremental contributions and intensifying the race for priority. 
This may also strongly encourage researchers to submit preprints to claim priority.

Researchers have tried to understand the computing communities, including the CSCW community \cite{wallace2017technologies}.
Correia et al. did some early bibliometric studies on collaborative scientific research in CSCW \cite{correia2019effect}.
Keegan et al. revealed that ideas introduced by elite researchers are advantaged over those introduced by newcomers in CSCW \cite{keegan2013structure}.
Our study extended this line of research by inspecting the community aspects of preprinting practices through talking to academics in AI and HCI.

\section{Methodology}

Below, we elaborate on different components of the IRB-approved study, including participant recruitment, semi-structured interviews, and data analysis.

\subsection{Participant Recruitment}

As our goal was to understand how academics perceived and practiced preprinting and how they adapted to the accelerating computing fields, we purposely recruited graduate students and professors in Computer Science and Information Sciences, whose research interests were in AI, HCI, or the intersection of the two fields.
We recruited through both social media (X) posting and direct emailing, as only a few participants were identified through recruitment on X.

The recruited fifteen participants have diverse backgrounds.
Four participants self-identified as HCI researchers; eight participants self-identified as AI researchers; three participants self-identified as scholars working at the intersection of HCI and AI.
They included 1 master's student, 7 PhD students, 1 postdoc, 3 assistant professors, 1 associate professor,  1 full professor, and 1 software engineer.
They were equally distributed between male (N=8) and female (N=7).
Their research experience ranged from 2 years (P13 and P14) to 30+ years (P2).

The participants had diverse, often interdisciplinary, research interests.
For example, P9 self-identified as an HCI researcher, but often worked at the intersection of HCI and AI (a.k.a. human-centered AI).
He aspired to use AI to automate designers’ work.
P8 is another example who explored the interaction between machine learning and law and focused on social media regulation.
P5 was interested in the governance of sociotechnical systems, including AI systems.
There were also technical AI researchers, such as P14 and P15, who worked on causal inference, and P1, who worked on NLP and embodied AI.

More information about the participants can be found in Table \ref{demographis:tab}.


\begin{table*}[h]
\centering
\begin{tabular}{@{}lllll@{}}
\toprule
\textbf{Number} & \textbf{Gender} & \textbf{Profession} & \textbf{Years of research experience} & \textbf{Research field(s)} \\ \midrule
P1  & M & PhD Student         & 5   & AI      \\
P2  & F & Professor           & 30+ & AI      \\
P3  & M & PhD Student         & 6   & AI, HCI \\
P4  & F & Postdoc             & 10  & AI      \\
P5  & F & Assistant Professor & 19  & AI      \\
P6  & F & Assistant Professor & 10  & HCI     \\
P7  & F & Associate Professor & 11  & HCI     \\
P8  & M & PhD Student         & 7   & AI      \\
P9  & M & PhD Student         & 3   & HCI, AI \\
P10 & M & Assistant Professor & 7   & HCI, AI \\
P11 & F & PhD Student         & 5   & HCI     \\
P12 & M & Master's Student      & 3   & HCI     \\
P13 & M & Software Engineer   & 2   & AI      \\
P14 & M & PhD Student         & 2   & AI      \\
P15 & F & PhD Student         & 5   & AI      \\ \bottomrule
\end{tabular}
\vspace{6pt}
\caption{Demographic information of the participants (N=15).}
\label{demographis:tab}
\end{table*}

\subsection{Semi-structured Interviews}

For the semi-structured interviews, we started by asking about the participants' personal background, including research area and level of research experience.
This part was used to build rapport with the participants, who tended to enthusiastically introduce their own research.

We then delved in-depth, asking about the participants' perspectives on preprinting, regardless of whether they had practiced it.
For those who had practiced preprinting, we asked about their preferred preprinting practice, e.g., at what stage of the publication lifecycle did they preprint, what proportion of their papers was preprinted, their initial preprinting practices, changes in preprinting practices (both at personal and community levels), motivations for preprinting, overall experience and perception of preprinitng, and perceived quality of preprints.
For those who had not practiced preprinting, we still asked about their experiences and perceptions of preprinting, perceived quality of preprints, and observed changes in preprinting practices in the community.
We additionally asked them about their motivation for the non-use of preprinting and whether they would preprint in the future.

In the end, we asked questions related to the interaction between preprinting and computing culture, including the advent of ``pre-doc'' programs, fast tech development, the growth of submission numbers in conferences, and the growth of people working in computing fields.
These characteristics reflected the fast-production nature of the computing disciplines.

The full interview script is attached in Appendix \ref{appendix:script}.

We audio-recorded the interviews, which lasted around one hour, with participants' consent.
The recordings were later transcribed using Otter.ai, an AI-based transcription tool, and errors were manually corrected.

\subsection{Data Analysis}

We adopted a thematic analysis \cite{clarke2017thematic} approach to analyze the interview transcripts.
Two authors collaboratively analyzed the data and reached a consensus.
We first exploratorily examined the transcripts and conducted open coding.
Then, emerging themes and sub-themes were organized into a hierarchical structure using XMind, a mind-mapping tool.

The first major theme concerned preprinting practices of the participants and the community. 
Under this theme, we found rich discussions regarding the evolving preprinting practices.
For example, through their career trajectory, researchers tended to preprint more often and earlier in the research lifecycle (e.g., post-acceptance to post-submission).

The second major theme concerned researchers' motivations for preprinting, rooted in the fast-production computing culture.
The motivations included increased visibility among numerous papers, community norms, fear of scooping in a fast-production field, career goals, and flaws in the current publication system.

The third major theme concerned the disadvantages of preprinting.
Some disadvantages included the lack of recognition for preprints, potential violation of anonymity/biased reviews, quality control issues, and the risk of scooping.

The fourth major theme concerned the comparison between preprints and publications in terms of quality and novelty, with diverse judgments from the participants.

In this paper, we use anonymized quotes to illustrate our findings.

\section{Results}

\subsection{Diverse and Evolving Preprinting Practices}

\subsubsection{Diverse Preprinting Practices}

The participants exhibited a wide range of preprinting practices. 
P4, P8, P10, P12, and P14 consistently made all their papers available on arXiv or other preprint servers. 
In contrast, P13 and P15 had never preprinted any of their work. 
Other participants fell somewhere in between, varying in the proportion of their work they chose to preprint. 
Despite these differences, most participants agreed that the rise of preprinting is primarily driven by the pressure to publish in today’s fast-paced research environment. 

\subsubsection{The Rise of Preprinting}

Many participants discussed the rise of preprinting with more uniformity, as several explained it was driven by the pressure to make their work publicly available. P10 explained: 
\begin{quote}
    \textit{``I don't know how to recalibrate after everything that has been happening with machine learning. 
    And ChatGPT and the volume at which they are putting papers on arXiv, which is, you know, again, there are lots of factors at play. 
    Especially now that I will be starting as an assistant professor, I know that there are a lot of politics involved as well. 
    With annual reviews and you on your tenure track, you do want to get your paper out somewhere, whether it's arXiv or not. 
    You start stacking up the citation.''}
\end{quote}


Due to the common pressure to share one’s work publicly, preprinting has become more of a norm, as discussed by P7: 
\begin{quote}
    \textit{``It seems much more common and much more like the norm to just post on arXiv. 
    It's accepted and it's like an assumed thing.''}
\end{quote}

\subsubsection{Earlier Preprinting in the Publication Lifecycle}


Many participants like P8 discussed a shift in their preprinting practices, noting that while they used to wait for acceptance before uploading a paper to a preprint server, many conferences have now removed those restrictions, allowing authors to post their work at any time:
\begin{quote}
    \textit{``I would say, like a year or two ago, it would be like, we would archive, maybe once the paper is accepted, 
    then you work on the paper and archive it. 
    But now, you know, so my field is NLP, so our conference is ACL, and recently they got rid of the requirement. 
    It used to be, like, before you submitted a paper, you needed to be anonymous for two months, right? 
    Anonymous as in, no preprinting, no tweeting, and they got rid of this.''}
\end{quote}

\subsection{How Computing Culture Influenced Preprinting Practices}

The participants' motivations for preprinting were deeply rooted in the computing culture, which was characterized by fast production, huge volume of papers, competitive career advancement, and a flawed publication system.

\subsubsection{Preprinting as a Community Norm}

Nine participants deemed preprinting as a community norm in computing disciplines, particularly fields like AI and HCI.
P5 noticed that her postdoc advisor in CS ``preprinted everything.''
P8 referred to preprinting as a reflection of ``machine learning culture.''
P11 naturally learned and practiced the model of preprinting from her advisor, without even a formal conversation.

Another factor contributing to the rise of preprinting is the fear of missing out (FOMO). 
P1 noted that it is difficult to resist these feelings when social media constantly promotes the preprinting process, creating pressure for researchers to share their work as soon as possible to avoid falling behind the trend:
\begin{quote}
    \textit{``I mean, it's fast. 
    Sometimes you feel tired, and sometimes you have to work seriously to fight against FOMO, like fear of missing out, right? 
    People are aware that there is a lot of hypes on social media, like Twitter, or, you know, some news articles in WeChat or Medium or MIT Review, right? 
    It really takes some effort and mental health to get used to this kind of pace, and it's actually tiring, like, in the past, we could take our time, spend one year on a good problem, and really define it and work it, make it solid, and submit. 
    But now we sort of have to look for some low-hanging fruit. 
    You should either choose to give them up or do it fast." }
\end{quote}

Other disciplines did not always approve of preprinting.
P4's former advisor in biotechnology did not allow her to put her master’s thesis on a preprint server because of community norms and concerns about the intellectual rights of the institution.
P12 described how his previous advisor evaluated the value of preprinting,
\begin{quote}
    \textit{``His background is more in chemistry, so I think that's where he gets the idea that preprints are not valuable... 
    There's more value in the prestige of being published in a peer-reviewed venue. 
    So people care more about having a peer-reviewed publication rather than having it out there and accessible.''}
\end{quote}

Some scholars, like P5, adjusted their perceptions and practices with preprinting over time. 
P5 mentioned that computer science researchers are pushing to make preprinting the norm, and because she collaborated with them, she followed suit. 

\subsubsection{Preprinting for Better Visibility}

Thirteen participants cited visibility as an essential consideration when making preprinting choices, especially given the large volume of papers in AI and HCI conferences nowadays.
There is a consensus that preprints obtained as much traffic as formal publications, sometimes even more.

Papers tended to be buried and unnoticed, with the huge volume of papers in AI and HCI conferences.
P9 described how preprinting helped make his papers stand out from other papers in the same conference proceedings,
\begin{quote}
    \textit{``During conferences, a whole bunch of papers came out at the same time. 
    People are just going to get lost when they're looking at a whole lot of papers. 
    So it's probably a good strategy to, like, timeline-wise, differentiate your paper from all the other papers that get presented at the venue, especially before the conference. After the conference, people are just gonna get tired. 
    They probably don't want to [read papers], and it feels a little outdated as well. 
    So, before the conference, using a preprint to advertise your paper is a good strategy.''}
\end{quote}

For people like P14 and P15, preprinting helped them share ideas, get comments/feedback, and facilitate discussion.
P10 noticed that after he started to arXiv papers, the community’s interest in his research increased, and more people
attended his paper sessions at conferences.
He thought preprinting was a great approach to share ideas and
build influence quickly, ultimately helping researchers to
``carve out their own niche and identity in the field.''
Similarly, P8 thought preprinting helped boost the visibility of researchers through an official record of publication to date.
P7 described how preprinting helped her research get noticed and eventually led to the successful recruitment of a PhD student,
\begin{quote}
    \textit{``Getting more citations earlier is one, getting your research noticed is another.
    Like I mentioned, one of my articles is on arXiv, and one of my incoming PhD students has already read it and talked to me about it [during the interview].'' }
\end{quote}

To many (P3, P6, P7, P9, P10, P11, P12), preprinting is a way to get citations as a form of recognition.
P6 thought preprinting worked by creating ``citable objects'' for researchers. 
P3 illustrated how the non-use of preprinting would lead to loss of citations,
\begin{quote}
    \textit{``In some fast-moving fields, every year that you get delayed, you will lose a lot of citations.
    People are writing papers, so they need somebody to cite as evidence.
    So if you aren't there, there's somebody else, and you will not get those people citing you. 
    After some time, you will never get the recognition for your work. 
    And so arXiv, basically, is kind of like an insurance.''}
\end{quote}
Preprinting has succeeded in boosting citations in the participants' experiences.
For example, P11 acknowledged that her most cited paper was a preprint of a workshop paper, instead of papers published in prestigious conferences or journals.
She initially arXived this paper because the workshop did not have a proceeding and because she wanted to ``get collaborators for future work and signal the research to the community'' by
``putting things out there.''

\subsubsection{Preprinting for Career Goals}

Some participants like P14 practiced preprinting to gain an advantage in PhD applications or job applications, which are highly competitive nowadays.

Preprinting helped people realize their career goals.
P3 realized that tenure-track faculty positions are becoming harder to get in recent years, that visible research and work are important when looking for a job or applying for a grant, and that disseminating work is made easier with preprints.
P10 similarly believed preprinting is suitable for early career researchers, especially when applying for grants or fellowships.
P5 deemed the long waiting period between acceptance and being online unacceptable when she was on the job market.
P1 elaborated on how in-submission, preprinted papers added weight to job applications,
\begin{quote}
    \textit{``Nowadays, when we are looking for internships or full-time jobs in the industry, your employer will just look through your resume. 
    Even if some work is still in submission, they are still interested in the content. 
    So if you don't archive the paper, it can be pretty hard to share your research or demonstrate your expertise to your potential employers.
    That might be an issue, especially for beginner researchers.
    For senior researchers, probably fine, but for beginners,  even one paper really matters.''}
\end{quote}

\subsubsection{Fear of Scooping in a Fast-production Field}

Thirteen participants discussed the fast production culture in computing disciplines, with an exploding number of submissions in conferences, an increasing number of researchers in AI/HCI, and the fast evolution of technologies, particularly generative AI.  

Researchers like P9 and P14 were not against the trend of more people starting to work on critical technological problems. Still, they pointed out that the increasing attention on AI/HCI caused more pressure on the peer review system.
They simultaneously realized the importance of claiming novelty in competitive fields.

On the contrary, researchers like P12 were worried about the degradation of the quality of research after people flooded into the area without qualified expertise. 
He commented, 
\begin{quote}
    \textit{``People are just submitting more work, which makes papers less valuable.
    Now, there are people who are not in PhD programs that are publishing, such as undergrad students. 
    You need three or more papers from top-tier conferences to get a job.''}
\end{quote}
P8 believed the surge of conference submissions has been ``breaking the system.'' 
P1 felt tired and acknowledged that he had to ``work seriously to fight against the fear of missing out.''
P11 missed the earlier part of her PhD years, where there was less trend chasing. 
Nowadays, ``it’s hard to stay on top of everything.''

In such a fast-production culture, state-of-the-art research is constantly evolving, and it is increasingly difficult to claim novelty.
P8 recalled that two arXiv papers, which were time-stamped two days apart, were highly similar.
Either paper would have been scooped if they were not preprinted.
He commented that since everything comes down to deep learning, people are thinking about similar things all the time.
P1 decided to practice preprinting for the same reason,
\begin{quote}
    \textit{``The pace of AI research has been pretty crazy. 
    If you don't arXiv your paper on time, and you wait for the paper to be accepted, it might take one year or more. 
    If you wait for that amount of time, usually, other researchers may scoop your idea. 
    That's also my concern. 
    Like, your research has to be on time because the pace is pretty fast nowadays, right?''}
\end{quote}

Preprinting became a solution for researchers like P1 and P10, who treated it as a timestamping tool.
P1 explained his rationale for timely preprinting: 
\begin{quote}
    \textit{``When you work on a problem, it is usually a problem in time, but when you wait for a few months and then you release a preprint of that paper, in many cases, either the problem is no longer interesting, or the state of the art has changed.''}
\end{quote}
P14 similarly used preprinting to prevent ideas from being scooped by ``putting a time stamp on the idea.''

\subsubsection{Flawed Publication System}

Some motivations for preprinting are related to intrinsic drawbacks of the publication systems, including prolonged review time, unprofessional/mean reviewers, and paywalling.

Conferences and journals often failed to make papers visible in a timely manner and catch up with the fast pace of AI/HCI research.
P3 noted that conference and journal publication took months or even years.
If rejected, there will not be proof that the team does the work -- for him, preprinting becomes a necessity.
He elaborated,
\begin{quote}
    \textit{``Even though the field is advancing, the conferences took months to give us feedback. 
    And so what happened is we are afraid that our work will lose novelty, so putting it on arXiv seems like a better option.''}
\end{quote} 

Delays can occur at different stages of the publication process.
The delay between acceptance and publication was mentioned by P5, whose article took almost two years to appear online after acceptance, due to the backlog of COVID-time peer review.
She deemed it essential to preprint forthcoming papers, especially for pre-tenure academics.

Seven participants complained about the long turnaround in the publication process caused by peer review.
P9 and P11 tended to arXiv their CSCW papers, as the review process took considerably longer than other HCI conferences, potentially making their research less timely by the time of formal publication.
P3, who used to believe the peer review process is the gold standard for validating scientific research, now thinks there are significant problems with it, particularly its role in making research ``stuck for years'' instead of being presented to the public in a timely manner.
One of P7's papers was rejected after minor revision, and the whole peer review process took two years.
P5 noticed that the peer review process got even slower after COVID, which resulted in her increasing preprinting practice.
The participants commonly used preprinting to combat the slow peer review process.
P7 provided an example, 
\begin{quote}
    \textit{``So I see more advantages [of preprinting] now with the slow review process of getting something out there.
    [With preprinting] My article isn't sitting there for two years, and then I'm trying to publish it somewhere else, and it's two years old. 
    If I had done something like that [preprinting], at least, you know, it could have gotten the ideas out there first.''}
\end{quote}

Eleven participants complained about the quality and over-strictness of peer review, which motivated them to practice preprinting more and earlier in the publication lifecycle.
P7 realized how challenging it is to get papers accepted into competitive conferences -- they have to be ``in that niche of what exactly it is that they [reviewers] want.''
P5 pointed out that gatekeeping is implemented by design in CS conferences -- the process makes the conferences have a lower acceptance rate and be perceived as more prestigious by the community.
Publishing in new or interdisciplinary fields is even more challenging.
P8 found it hard to publish in the emerging, interdisciplinary area of AI and law.
Reviewers from law or AI backgrounds often suggest that he submit his work to venues in the other discipline.
He further found it hard to publish non-technical pieces, such as 3-page policy briefs in conferences, so he turned to arXiv and his own blog for scientific communication.
P8 observed that since industrial labs invested a lot in research, they could not stand being rejected by conferences and losing the novelty of their research.

Peer review hindered innovation, according to some participants.
P3 believed that the current peer review system suppresses novelty in that reviewers tend to reject papers that are too novel or of too broad a scope.
P12 commented,
\begin{quote}
    \textit{``Famous papers did not get accepted to conferences, but preprints became the actual papers.''}
\end{quote}
P8 similarly speculated that the classical, famous AI papers might not be accepted by reviewers nowadays, who only read some parts of the papers.

The participants frequently encountered low-quality reviews.
P9 recalled how reviewers who were not domain experts asked him to cite irrelevant work for the literature review.
P8 found that most of the time, reviews were not helpful, did not contain much detail, and were short, and indicated that the reviewers did not spend much time reading the papers.
P5 recalled frustrating gatekeeping experiences where quantitative researchers reviewed her qualitative work, and no constructive feedback was provided.
She rarely experienced cases where all three reviews were reasonable.
P15 cautioned that the use of AI might be a catalyst for the further decline in review quality.

P7 was frustrated by the fact that even if reviews were not reasonable, authors were compelled to accommodate them, otherwise the papers would not be accepted,
\begin{quote}
    \textit{``We need to accommodate all the comments from the reviewers. 
    Some people mentioned that they think the reviewers' comments don't really make sense, but they have to accommodate the changes when they do the major revision. After the acceptance, they have a different version. 
    Their original version is on arXiv, and so that's the version they actually like the most.'' }
\end{quote}

P1 gradually lost trust in the role of peer review in determining the quality of research,
\begin{quote}
    \textit{``Nowadays, a lot of reviewing systems are just collapsing.
    They're not really giving you any feedback, or they're not technically replying.
    From time to time, you feel like you don't really have to wait for this kind of random process to set a milestone so that you can really publicize your work.''}
\end{quote}

P6 would preprint work that she deemed interesting, even if it has a possibility of being rejected by reviewers,
\begin{quote}
    \textit{``It [Preprinting] allows you to, you know, get work out into the conversation that you feel good enough about that even if it's rejected, you think is still worthy of being part of the conversation, especially when that conversation is happening in a really timely manner.    
    And you know, you won't find out for a couple of months whether or not the paper has been accepted to the conference.''}
\end{quote} 

Paywalling is another major hurdle for the dissemination of academic work.
P5 and P7 thought preprinting was important for equity, given the prevalent paywalls set by publishers. 
P5 noted that she even had to pay for her own articles, and that people outside of academia, such as lawmakers, often do not have a subscription like that in research institutions.
P4 did not have enough funding to pay for open access, but she still wanted to make her work as accessible as she could -- preprinting became the only solution for her.

\subsection{Drawbacks of Preprinting}

The participants expressed concerns about preprinting in addition to its advantages. 
Some of the surface-level drawbacks of preprinting include the outdated arXiv website, as mentioned by P8. 
The study further identified three major drawbacks of preprinting, namely, potential violations of anonymity policies, quality control issues, and lack of recognition from the community.

\subsubsection{Lack of Recognition and Validity}

Five of the participants highlighted what they viewed as a significant drawback of preprinting, i.e., the lack of formal recognition for preprinted work. 
They expressed concern that, despite the effort involved in producing and sharing preprints, these contributions are often not acknowledged in academic evaluations, tenure decisions, or grant reviews. 

As P2, P6, P7, and P8 pointed out, without the peer review process, it is difficult to validate the quality or credibility of a work.
Some participants preferred work that had undergone rigorous review.
P2 noticed that Google Scholar favored arXiv versions to formal publications, and would like her final publication to be discovered by others: 
\begin{quote}
    \textit{``So I, for one thing, because of the way arXiv fits into the ecosystem, once something's on arXiv, the actual published versions become harder to find. 
    So Google Scholar favors arXiv, right? 
    And I would prefer that when someone finds my papers that have been peer reviewed, they find them together with that information about peer review.''}
\end{quote}

Being asked to cite preprints annoyed some participants.
P6 talked about their frustration when asked to cite a preprint, which she believed had not been fully validated through the peer review process:
\begin{quote}
    \textit{``Why am I obligated to cite your preprint that would have been rejected nine times out of 10? 
    Does this meet a certain bar?''}
\end{quote}
P7 similarly questioned whether they should cite a preprint, as preprinting is not the same as formal publication, and she expressed concerns about the validity of preprints:
\begin{quote}
    \textit{``It's not the same as getting an actual publication. And sometimes I'm citing an article, and I'm like, should I cite it? If it's an arXiv, maybe not.''}
\end{quote}

P8 noted that preprints are not recognized as much as conference papers or journal articles due to the lack of peer review, but suggested that incorporating some form of review process into preprinting could be beneficial. 


\subsubsection{The Lack of Quality Control}

Aside from concerns about the validity of preprints, ten participants (P1, P2, P3, P4, P9, P10, P11, P12, P14, and P15)  raised concerns about the lack of quality control in preprinting. 
P9 and P14 noted that anyone could upload a paper to a preprint server, even if it was not of high quality.
They also pointed out that the ease of uploading papers to preprint servers contributed to an overwhelming volume of papers, making it difficult to filter out lower-quality work.
P1 stated,
\begin{quote}
    \textit{``It's just like a lot of papers with problems, like, it's not really peer-reviewed, like a lot of overclaiming, a lot of incomplete work in progress. We're just there.''}
\end{quote}

P3 discussed the quality issue of preprinting, which lacked the formal checks that traditionally help check for rigor. 
This process will allow faster dissemination of research, but it also increases the risk of spreading unverified or low-quality findings. 
He stated,
\begin{quote}
    \textit{``Overhaul will happen to preprinting and peer review probably. People say, `we’re being affected by the lack of peer review,' `peer review is not that scientific and transparent.' But an overhaul won’t happen without economic incentives.''}
\end{quote}

Another reason many preprint servers faced quality control issues was that people often used them to upload papers that had been repeatedly rejected elsewhere and were looking for a place to share them. 
P2 shared this sentiment:
\begin{quote}
    \textit{``I think there's a lot of garbage on arXiv. 
    Sometimes people use it for, like, here's a paper, I need to put it somewhere. 
    It's not getting accepted, or, for whatever reason, they're not submitting it somewhere.''}
\end{quote}


\subsubsection{Violation of Anonymity Policies and Reviewer Bias}

The last major drawback discussed by the participants was the potential violation of anonymity policies and reviewer bias, e.g., by revealing the authors’ identities (P9, P10) 

P1 talked about concerns that preprinting might unintentionally break the anonymity rules and that this could contribute to visibility or fairness issues, especially for less prestigious institutions, 
\begin{quote}
    \textit{``I mean, Michigan is a pretty top school, but compared to Stanford, you know, those people receive even more attention. 
    Some relatively less privileged research groups might suffer from more visibility issues and fairness issues, right? 
    I do believe there is a fairness issue.
    People tend to cite and read papers from famous research groups and from famous scientists.''}
\end{quote}

P5 raised similar concerns about how preprints can potentially violate anonymity rules in the peer review process. 
She pointed out that during the review process, reviewers might search related work using keywords from the submission. 
In doing so, it is possible that they could come across the preprinted version of the paper, which may inadvertently reveal the authors' identity. 

P12 further pointed out that the use of social media could further exacerbate anonymity concerns, as preprints were made highly accessible. 
He noted how some authors got publicly exposed to scrutiny without undergoing peer review, and how this could lead to the exposure of authors' identities before or during the peer review process.

\subsection{Preprints vs. Publications}

The opinion on the quality/novelty of preprints vs. peer-reviewed publications is diverse among participants, with most expressing that neither is inherently superior. 
For example, the peer review process is often subjective, with reviewers offering inconsistent feedback, leading to potential delays or rejection of valuable research. 
While preprints offer quicker access to research, they lack the formal quality control of peer-reviewed publications. 


Some participants, like P1, P10, and P12, believed journal and conference publications were of higher quality than preprints because of the peer review process. 
P10 elaborated on this opinion: 
\begin{quote}
    \textit{``Conference papers are considered higher-quality than preprints because reviewers ensure quality. 
    You mentally prepare to spend a year getting papers out and accept that this is the reality in academia. 
    It's also a matter of patience and personality.''}
\end{quote}




Most of the participants (P1, P2, P3, P4, P5, P6, P14, P15) expressed that neither preprints nor conference/journal publications were of higher quality by default. 
Some participants, such as P14 and P15, expressed that there were very good preprints, and that journal/conference papers were not necessarily of higher quality, depending on the quality of the venues. 
P14 elaborated,
\begin{quote}
    \textit{``I don’t think preprints can be of high quality because they do not have any review. 
    But journals might not be high-quality either. 
    Some journals can just be paid to publish. 
    The best way is to ensure quality by yourself."}
\end{quote}


P5 argued that conferences could constrain researchers' creativity, as researchers often had to conform to the specific expectations or formats of the venue. 
P2 similarly suggested that the slow pace of the review process could hinder innovation, causing some papers to become outdated or less groundbreaking by the time they were published. 
Along these lines, P4 noted that reviewers may sometimes harm a paper through overly critical or misguided comments, which could negatively impact its overall quality.:
\begin{quote}
    \textit{``Reviewers sometimes ruined it (“do this, do that” “background too long”). 
    When you publish your research in a preprint, people can have the whole idea of what you want to say. 
    So in that, I think preprint is better, right?''}
\end{quote}






A few participants, such as P8 and P9, believed that peer review did not make a significant difference in the current mode of publication. 
P8, for example, explained that he was no longer interested in looking through conference proceedings because he now prioritized the relevance of papers to their needs, rather than whether they were formally peer-reviewed and published in a conference. 
This reflects the perception that peer review does not necessarily ensure the quality or relevance of the research papers.
\section{Discussion}

\subsection{Research Implications}

Past research has uncovered multiple benefits of preprinting in disciplines beyond computing, such as timely dissemination and visibility \cite{fraserMotivationsConcernsSelection2022}, providing a record of productivity for career-oriented purposes \cite{sarabipourValuePreprintsEarly2019}, sharing unfavorable results \cite{bourneTenSimpleRules2017}, preventing scooping \cite{chiarelliPreprintsScholarlyCommunication2019}, boosting citation \cite{fraserRelationshipBioRxivPreprints2020}, among others.
While some researchers argued that early arXiving does not provide an advantage for paper acceptance \cite{elazar2024estimating}, implicit benefits such as attracting potential students through preprinted papers were valued by participants who would like to advance their career and research.
Some participants began to arXiv papers earlier in the publication lifecycle, intrinsically practicing a two-stage advertisement of research: after project completion and after conference/journal endorsement. 

Our participants in the computing disciplines additionally expressed some other motivations for preprinting, such as following the community norm and bypassing the flawed publication system.
In particular, the delay of research communication can arise from the prolonged waiting time between acceptance and ``online,'' and low-quality, while overly-strict peer review.
Some academics started to question the fundamental peer review system: Why do authors have to address reviewers' comments? Why are reviewers' judgments necessarily more legitimate than authors'? 
Such thinking enabled some participants to decouple the correlation between peer review and paper quality and more naturally accept and practice preprinting.

Past research has also revealed drawbacks of preprinting, such as a lack of credibility \cite{adarkwahArePreprintsThreat2025}, quality control challenges \cite{smartEvolutionBenefitsChallenges2022}, misuse of illegitimate research findings \cite{niPreprintNotPreprint2023}, and violations of anonymization policies \cite{fatoneChallengesDoubleblindPeer2020}.  
Our study confirmed these drawbacks from the perspective of academics.
The drawbacks, particularly the ones related to quality control, can be further strengthened by the misuse of generative AI.
With the increasing use of LLMs, researchers, particularly junior researchers, may use these generative models to generate references, which can be totally fabricated (e.g., an arXiv link with a non-existent ID) and may hurt the reputation of a scholarly article significantly.
Some participants, however, noticed that the credibility of preprints has been rising in recent years, with the evidence that reviewers frequently ask authors to cite preprints.

\subsection{Design Implications and Challenges: Toward a Better Publication Ecosystem}

Computing scholars have difficulty keeping up with continual technological breakthroughs \cite{togeliusChooseYourWeapon2024}.
Echoing this research, our study revealed that not only scholars, but also conferences and journals have difficulty catching up with the fast pace of AI/HCI research and the exploding number of paper submissions.
Low-quality, even AI-generated reviews, and unqualified reviewers frustrated authors and significantly delayed scientific communication.
The research community should reconsider the role of AI reviewers \cite{wiechert2024will} in the peer review process, which may not necessarily perform more poorly than human reviewers, and may alleviate the pressure caused by the huge volume of submissions. 

The rise of medicine-related preprints \cite{gianolaCharacteristicsAcademicPublications2020} during COVID highlighted the importance of disseminating socially important research promptly.
The slow and often overly strict peer review system can hardly achieve this goal.
With the intended low acceptance rate at HCI and AI conferences, socially significant research on topics such as AI governance may not be adopted in regulatory practices in a timely manner.
According to Carson et al., the “publish-or-perish” culture and the low acceptance rates at prestigious CS conferences are factors that motivate researchers to claim priority through preprinting \cite{carsonOverCompetitivenessAcademiaLiterature2013}. 
With preprinting, our participants were able to make accessible research that prestigious conferences could potentially reject -- they were satisfied with such research and believed it could benefit society, no matter how the reviewers judged. 
P5 insightfully suggested alternative venues (e.g., research magazines) and forms (e.g., proposals) for communicating research of urgency, social importance, and practical and policy implications.

Some participants insightfully imagined new publication modes or behaviors.
For example, P8 proposed a decentralized scientific communication mechanism that emphasized building small social networks instead of focusing on popularity.
This way, academics who are not popular also receive attention.
P2, frustrated by the ``publish or perish'' culture, pointed out a more sustainable solution, i.e., working on topics that are less likely to be scooped.

\subsection{Limitations and Future Work}

There are limitations of our study.
First, our sample size is relatively small and is skewed toward PhD students.
Nevertheless, we used in-depth interviews to obtain first-hand insights from early-career academics, who have not yet fully learned the rules of the ``academic game'' and are often frustrated by the publication ecosystem.
Future research should follow up with a larger-scale survey study to obtain more diverse perceptions of preprinting in computing disciplines.

Second, the results reflected views from academics with strong opinions about preprinting and peer review.
It is equally interesting to understand how those without strong opinions about these academic aspects navigate the fast-production culture in computing disciplines, which is left out of the scope of the current study.

\section{Conclusion}

Through semi-structured interviews with 15 academics in HCI and AI, we obtained an understanding of their practices and perceptions of preprinting.
Diverse and evolving preprinting practices were revealed, suggesting a trend of earlier preprinting in the publication lifecycle.
Computing culture, characterized by the increasing number of papers and the accelerating pace of research, influences academics' decision to preprint.
Preprinting has the potential to decouple the historically close tie between peer review and paper quality/innovation.
We discuss potential solutions and challenges toward a better publication ecosystem, which is more open, prompt, and efficient.

\bibliographystyle{IEEEtranS}
\bibliography{references}

\appendix

\section{Interview Script}
\label{appendix:script}

\subsection{Interview Protocol}

\noindent \textbf{Background}

\noindent 1. Could you tell me a bit about yourself? \\
2. What research do you work on? \\
3. How many years have you been a researcher? \\

\noindent \textbf{Preprinting} 

\noindent 4. Have you uploaded your papers to arXiv or other preprint servers?
    \begin{itemize}
        \item If yes, ask all questions starting 5.
        \item If no, ask
        \begin{itemize}
            \item Was there an opportunity where you could have uploaded research to a preprint server?
            \item Why did you choose not to upload research to a preprint server?
            \item If given an opportunity, will you upload research to a preprint server? Why or why not?
            \item Do you know colleagues who have uploaded research to a preprint server?
            \item What do you think are their rationales for uploading research to a preprint server?
            \item Ask questions 10-19 
        \end{itemize}
    \end{itemize}
    
\noindent 5. When do you usually upload your papers to a preprint server? Pre-submission, post-submission, or post-acceptance/rejection?  \\
6. What proportion of your papers are on a preprint server?  \\
7. When did you upload your first preprint paper? Can you tell me about your rationale for preprinting back then? \\
8. Were there changes in your career? \\
9. What is your motivation for preprinting? \\
10. Are there any differences between you and your advisor(s)/advisee(s) regarding preprinting? \\
11. What is your overall experience with preprinting? \\
12. What is your overall perception of preprinting? Pros and cons? \\
13. Do you think preprints or journal/conference papers are more innovative/of higher quality? \\
14. Have you seen changes regarding preprinting in your field? 
\begin{itemize}
    \item Would you put research on a preprint server in the 90s?
\end{itemize}

\noindent \textbf{Relevant factors}

\noindent 15. How do you view the advent of “pre-doc” programs?  \\
16. How do you view the fast tech development in recent years? \\
17. How do you view the growth of submission numbers in computing, especially AI conferences?  \\
18. How do you view the growth of people working in computing? \\
19. How do you think these factors contribute to the rise of preprinting in computing? \\

\end{document}